# Optimal Productivity of Succoring Patients System using Mobile GIS Based on WCF Technology

**Ayad Ghany Ismaeel**
Department of Information System Engineering, Erbil Technical College, Hawler Polytechnic University (previous FTE- Erbil), Iraq. E-mail: dr_a_gh_i@yahoo.com
Alternative E-mail: dr.ayad.ghany.ismaeel@gmail.com

**Sanaa Enwaya Rizqo**
Computer Institute - Department of Network Administration
Duhok, Iraq.
Email: sana_ar2001@yahoo.com

*Abstract*—Depending on references of the World Health Organization there is large numbers of sick persons with different diseases worldwide, and without succor at a suitable time this could lead to fatality of the patients. This paper offers a succoring system controlled by the patient based on the patients' location. The proposed system is the first tracking system using mobile GIS based on WCF technology to offer online succoring (**24 hour a day**), but really works only when the patient sends request for succoring. The patients will send a request (SMS by click one button) contains his ID, Longitude and Latitude via GPRS network to a web server containing a database, which the patient was registered previously on it. Then the server will locate the patient on Google map and retrieve the patient's information from the database. This information will be used by the server to send succoring facility and notify the nearest and most suitable ESC; moreover, the server will send SMS over IP to inform the patient emergency contacts and emergency hospital. The optimal productivity for proposed succoring system appears in handling a large number of requests within short period at rate of one request/need succoring per sec as result of using mobile GIS based on WCF technology. Furthermore, the process of request and reply for emergency cases of the patients achieved in cost-effective way due to this technology, which allow sending data (SMS over IP) via Internet using GPRS network. The proposed system can be implemented in a minimum configuration (hardware and software) to minimize the overall cost of operation and manufacturing.

*Keywords- WCF; WPF; Build-in GPS; GPRS; Mobile GIS; SoIP; Tracking System; Google Maps API*

## I. INTRODUCTION

All According to the WHO, the numbers of patients with chronic/non-chronic diseases are increasing every year [1]. As an example, 300 million people affected by Asthma in 2010, and this number will goes up to 400 million people world-wide by 2025[2]. Another chronic disease is diabetes. Estimations provided by the WHO pointed out that in 2025 the diabetics will reach 333 million worldwide from 135 million patients in 1995[2]. Moreover, 1,000,000 babies are born with congenital heart disease, which is a birth-defect, each year [1].

These three important diseases give a view of the big problem worldwide without taking into consideration the other diseases. This problem become more effective if there is not who offer succoring and help at a suitable time for these patients, while they are in school, shopping or with their friends. Therefore, without a proper succoring system the fatality rate may increase among those patients [1].

There are several papers dealt with dedicated alarm systems designed for medical reasons. The systems presented in [7, 8] requested devoted microprocessor and I/O devices (screen and keyboard). Moreover, these systems mainly relied on the dedicated tracking technology to identify the location of the patient. These dedicated alert systems used Global Positioning System GPS to provide the tracking facility. The reviewed papers may argue that the use of such systems will reduce the overall cost. The traditional tracking systems developed so far are using a handheld GPS receiver device to pinpoint the location depend on real time tracking and continuity of the tracking interval [3].

Therefore, there is a needed for succoring system can track the patients only when they require succoring and there is no one around them aware of their condition. Time and the quality of providing this service are a key factor for these patients. Consequently, new techniques such as mobile GIS based on Windows Presentation Foundation WCF technology are used to provide a satisfactory system in achieving low cost and higher productivity.

Windows Presentation Foundation WPF is Microsoft's new generation User Interface UI framework to create applications with a rich user experience. It is part of .NET framework. WPF possesses rich interactive ability. It also provides the UI, 2D/3D graphics, documents and multimedia support [15]. One of the key concepts of WPF development is an almost total separation of design and functionality. This separation enables designers and C# developers to work together on projects with a degree of freedom that previously required advanced design concepts or third-party tools. This functionality is to be welcomed by all; small teams and hobbyist developers as well as huge teams of developers and designers that work together on large-scale projects [9]**.**WPF includes the familiar standard controls, which draw every text, border, and background fills itself. As a result, WPF can provide powerful features that provide alter way for any piece of screen content to be rendered. By using these features, it is possible to restyle common controls, such as buttons, without writing any code. Similarly, transformation objects can be used to rotate, stretch, scale, and skew anything in the user interface. Beside, WPF's baked-in animation system can be used to do it right before the user's eyes [10].

The WCF is an extension of the .NET framework to build and run connected systems. It is also a flagship product of Microsoft on service oriented architecture SOAP, which is appeared for the first time in .net3.0 and supported the REST web service since .net3.5. WCF enables developers to build secure, reliable, transacted solutions that integrate across platforms and interoperate with existing investments. It allows application to make function as services to provide the client service requestor [11].

The architecture of WCF appears remarkably simple. The services are deployed, discovered and consumed as a collection of endpoints, each of which is the fusion of the Address (A), Binding (B) and contract(C), known as ABC. The service must have at least one endpoint, as illustrated Fig. 1 [12].

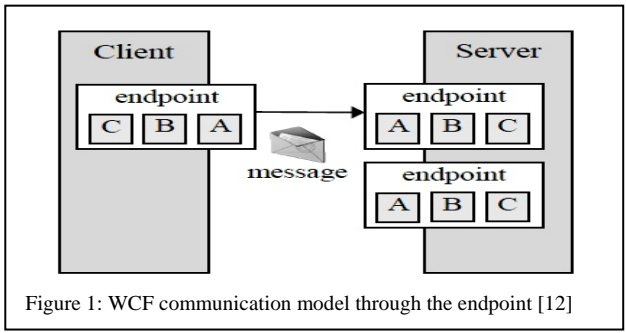

Figure 1: WCF communication model through the endpoint [12]

Throughout the next six paragraphs the various Message Exchange Patterns MEPs supported by WCF are presented [13]:

*A.  Request-Reply:*

Is WCF's default operation mode using this MEP, the client issues a call to the service in the form of a message and blocks until it gets a reply. If the service does not respond within the specified timeframe, the client will get a Timeout-Exception.

*B. One-Way:*

In case an operation has no return value and the client does not care about success or failure of the call, WCF offers one-way operations to support this kind of "fire-and-forget" invocation. Contrary to request-reply calls, one-way calls usually block the client only for the briefest moment required to dispatch the call. As only a request message but no reply message is generated by WCF, values (as well as exceptions thrown on the service side) can't be returned to the client.

*C. Callback/Duplex:*

Using duplex communication (or callbacks) WCF allows a service to call back to its clients and invoke a client method. Callback operations are especially useful when it comes to events and notifying the client that some event has happened on the service side. However, in order to enable the service to call back to the client, it has to know the client's endpoint. Therefore, it is necessary for the client to call a service method Fig. 2; step 1 first, which saves the callback channel to the client's endpoint for later use Fig. 2; step 2. Through that channel it is possible for the service to send messages to the client and invoke certain methods.

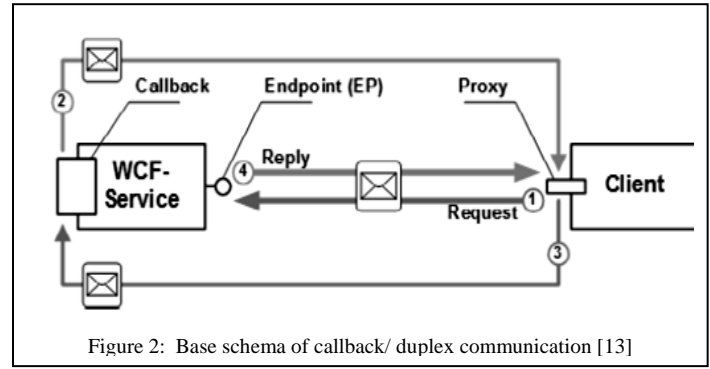

Figure 2: Base schema of callback/ duplex communication [13]

*D. Streaming:*

When client and service exchange messages, they are buffered on the receiving side and delivered only once the entire message has been received. This means, that the client is unblocked only if the request message and the service's reply message have been sent and received in its entirety. This works well for small messages. However, when it comes to larger messages (e.g. multimedia content) blocking until the entire message has been received may be impractical. Therefore, WCF enables the receiving side to start processing the receiving data while the message is still being received. This type of operation is called streaming transfer mode and improves throughput and responsiveness with large payloads.

*E. Asynchronous Calls:*

Using asynchronous calls the client will not block and control returns immediately after the request has been issued. The service then executes the operation in the background. As soon as the operation completed execution the client is provided with the results of the execution. Asynchronous calls improve client responsiveness and availability.

*F. Queued Calls*:

Queued messages use Microsoft Message Queue MSMQ. WCF encapsulates each SOAP message in an MSMQ message and posts it to the designated queue. Thus, the client does not communicate directly with a certain service but with a message queue. As a result, all calls are inherently asynchronous and disconnected. On the service side, queued messages are detected by the queue's listener, which sequentially takes messages from the queue and dispatches it to a service instance [13].

WCF fuses different distributed technologies and unifies them under one programming model, as shown in Table 1. In addition, WCF is designed to reduce development complexity and make developers that are already familiar with existing technologies such as ASP.NET, .NET remoting and Web Services feel comfortable with the new technology. Besides, it provides integration and interoperability with the existing .NET technologies [12].

TABLE I. TYPE STYLES [12]

|  | ASMX | .NET Remoting | Enterprise Services | WSE | System Messaging | System Net | WCF |
|---|---|---|---|---|---|---|---|
| Interoperable Web Services | yes |  |  |  |  |  | yes |
| .NET to .NET Communication |  | yes |  |  |  |  | yes |
| Distributed Transactions |  |  | yes |  |  |  | yes |
| WS-* support |  |  |  | yes |  |  | yes |
| Queued Messaging |  |  |  |  | yes |  | yes |
| RESTful Communication |  |  |  |  |  | yes | yes |

## II. RELATED WORK

Ayad Ghany Ismaeel [2012] suggested an emergency system for succoring sick child locally when he required that, and there isn't someone knows his disease. In this emergency system the child will send SMS contains his ID and coordinates (Longitude and Latitude) via General Packet Radio Service GPRS network to the web server, in this step the server will locate the sick child on Google map and retrieve the child's information from the database which saved this information in registration stage, and base on these information will send succoring facility and at the same time informing the hospital, his parents, doctor, etc. about that emergency case of the child using the SMS mode through GPRS network again [1].

Ruchika G. and Reddy B. [2011], proposed a cost effective method for tracking a human's mobility using two technologies via GPRS and GPS allowing the user's mobility to be tracked using a mobile phone which equipped with an internal GPS receiver and a GPRS transmitter. A mobile phone application had been developed and deployed on an android phone whose responsibility was to track the GPS location and send it to a remote location by creating a GPRS packet. The mobile's Irrational Mobile Equipment Identity IMEI number had been used as a unique identifier, which would be sent along with the coordinates. The person's position is further saved in a Mobile Object Database (MOD). The data would be first transferred into an XML file which will be fed as an input to a web application. This application was developed with JavaScript Ajax based Google Map API integrated into it which will be responsible for the showing the current location of the mobile phone. An inbuilt GPS receiver within a mobile phone used here, and GPRS used rather than using Short Message Service SMS for communicating the information to the server [5].

Khondker Shajadul Hasan, and etal. [2009], proposed and implemented a low cost object tracking system using GPS and GPRS. The system allowed a user to view the present and the past positions recorded of a target object on Google map through the internet. The object's position data were stored in the database for live and past tracking. This technique provided the advantage of the low-cost from the previous technique that depended on SMS for the communication to the server which turned out to be expensive. The system was very useful for car theft situations (alarm alert, engine starting, localizing), for adolescent drivers being watched and monitored by parents (speed limit exceeding, leaving a specific area), as well as for human and pet tracking [6].

The whole systems allow the user's mobility to be tracked using dedicated GPS or mobile phone which are equipped with an internal GPS receiver and a GPRS transmitter, i.e. most of the applications developed so far use a handheld GPS receiver device for tracking the location [6], and real time tracking and continuity on the interval of tracking may be very high cost specially when the server, and ISP are busy in the interval of tracking [4].

To overcome the problems above and to enhance the performance, throughput, of this type of system, the motivation is reached to system for succoring patients can achieve easy registration (from their homes) can select the nearest Emergency Service Center ESC when the patient need a succoring in emergency cases wherever her/his location the scouring facility (car, helicopter, boat life) can reach to her/his and in the type which is needed because the succor for blood pressure different from diabetes, heart, asthma, etc, so the proposed system will involve the following characteristics techniques and modes [1, 18]:

*A. Mobile GIS:*

Mobile GIS is the expansion of GIS technology from the office into the field. A mobile GIS enables field-based personnel to capture, store, update, manipulate, analyze, and display geographic information. It is an integrated software/hardware framework for the access of geospatial data and services through mobile devices via wireless networks. Mobile GIS integrates one or more of the following technologies (mobile devices, GPS) and wireless communications for Internet GIS access) as shown in Fig. 3 [14, 17].

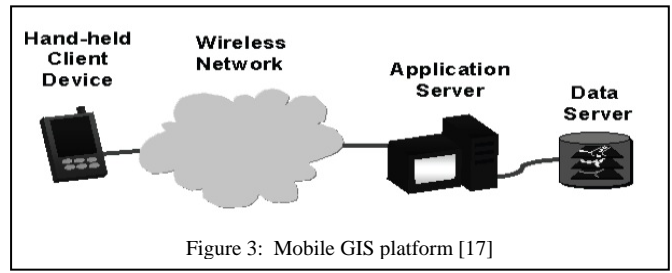

Figure 3: Mobile GIS platform [17]

The proposed system will base on supporting of a mobile build-in GPS technique on device Windows phone (OS 7.1).

*B. Modes of transmission as follow:*

1) *GPRS mode:* The mobile terminal sends the extracted data from GPS satellites through GPRS data channel to a special TCP/IP server (a PC with fixed IP address) linked to the Internet [1].
2) *SMS over IP SoIP mode:* is performed by connecting to the Short Message Service Center SMSC though a gateway as shows in Fig 4, which will use protocols, such as SMPP and SS7, to encode and decode the messages [16].

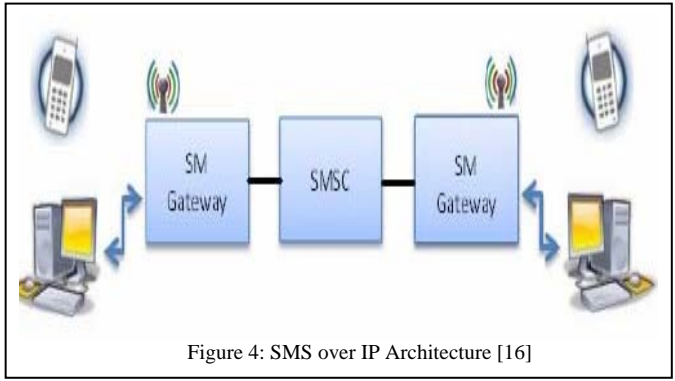

Figure 4: SMS over IP Architecture [16]

*C. Server of TCP/IP mode:*

The IP address for the server must be static. Therefore, the user will have to configure the IP address setting of the mobile terminal to align it to the server [1, 17].

*D. Windows Communication Foundation:*

WCF is a framework for building service-oriented applications. WCF allows the sending of data as asynchronous messages from one service endpoint to another [8].

III. PROPOSED SUCCORING SYSTEM BASED ON WCF

The architecture of suggested tracking system for succoring patients involves multiple stages, as shown in Fig. 5.

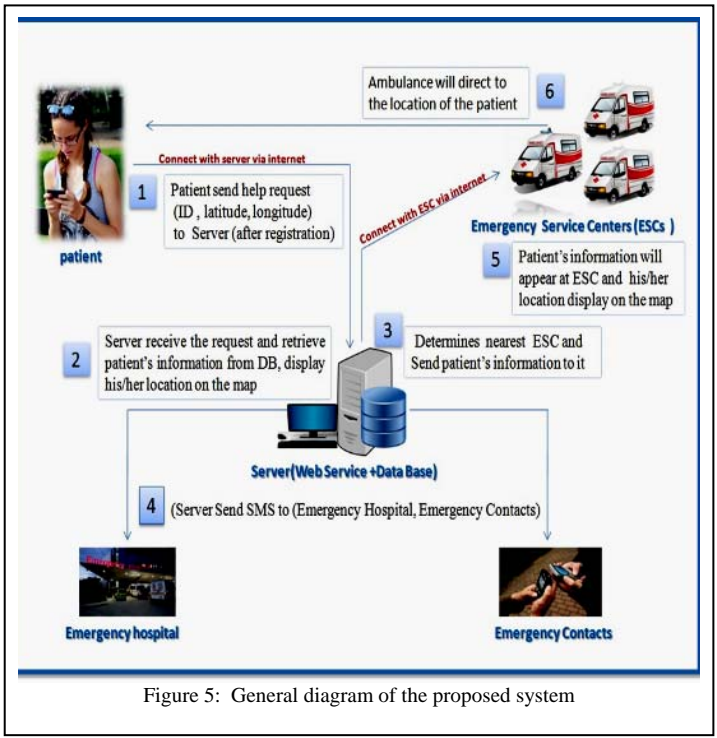

Figure 5: General diagram of the proposed system

The general tasks of proposed system summarized as flowchart shown in Fig. 6.

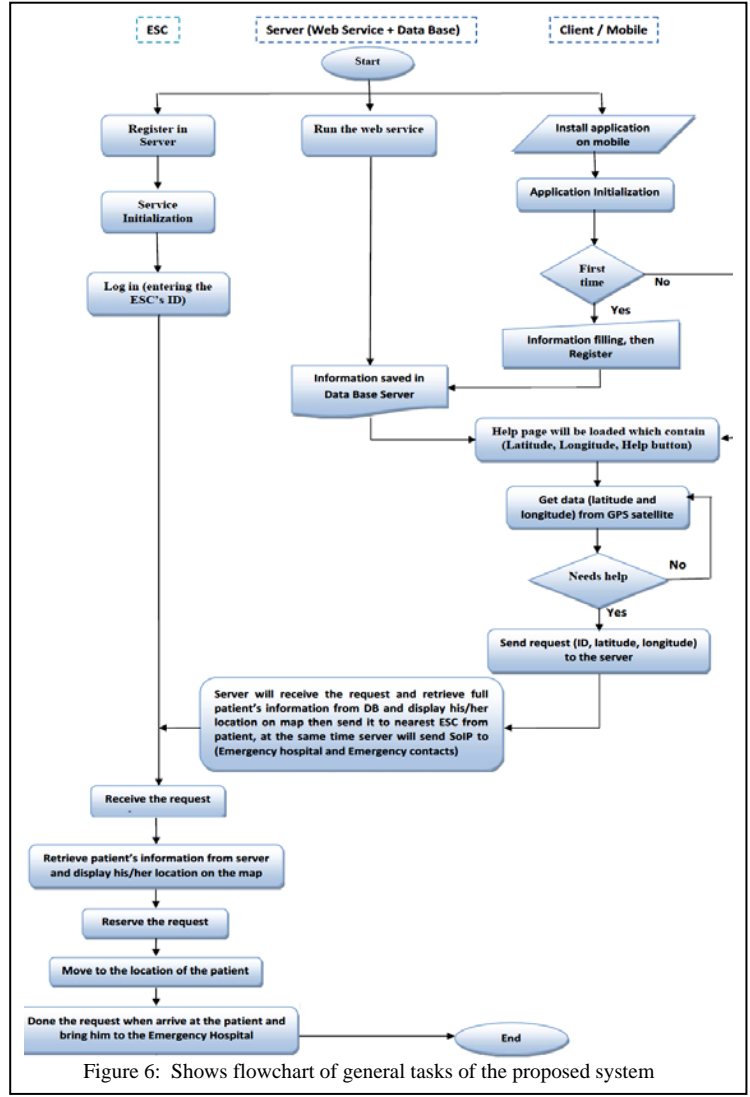

Figure 6: Shows flowchart of general tasks of the proposed system

The proposed system of tracking system for succoring patients using mobile GIS based on WCF technology must be consisting of three parts:

*A. The Client Side:*

In this system the client is a mobile hold by the patient. The windows phone (OS 7.1) with build in GPS receiver and GPRS transmitter work over GSM network is used in this system because it is more friendly with Microsoft packages, this compatibility will avoid the conflict in dealing with the software, and it is cheaper comparing to other smart phones (IPhone, Android, Blackberry) and other tracking devices (which can support GPS/GSM/GPRS technologies). The mobile application was developed with C# language in Visual Studio 2010 by using Silverlight for windows phone and windows phone SDK 7.1 technologies. The main steps of running mobile application are illustrated in Fig. 7, which shows flowchart of the main steps of running the client/mobile application under proposed system.

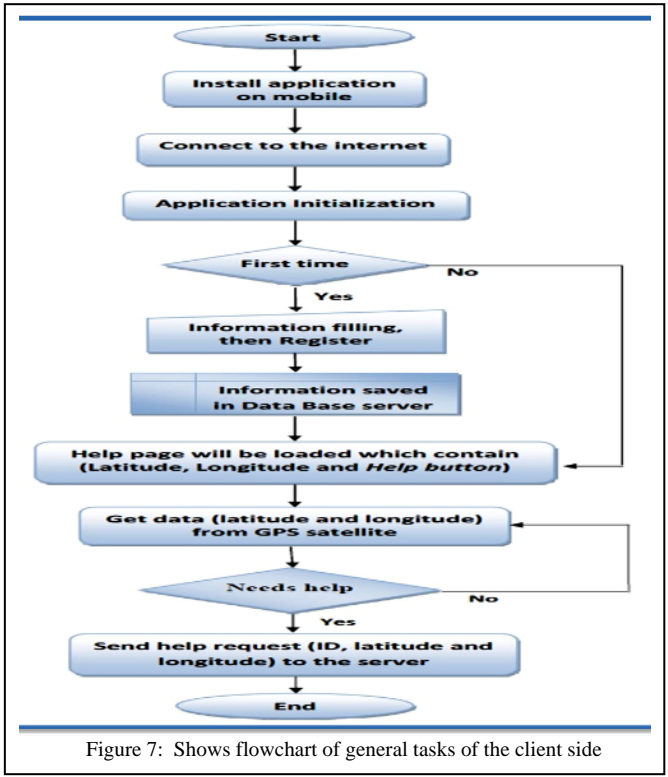

Figure 7: Shows flowchart of general tasks of the client side

### B. The Server Side:

The server is a PC provided with an internet connection and has static IP address, supported by Windows operating system, Microsoft .NET framework 4.0, and Microsoft SQL Server 2008 R2, which will be used for saving the incoming data from mobile. The server should be running 24 hours a day, but provide the tracking service to the patient only when he/she sends the help request to the server not all the time in day. The server works automatically when receiving the request from the client and sends it to the nearest ESC also send the SMS to the emergency contacts of patient (client) and to emergency hospital. The server uses algorithm of Haversine equation, which is an equation important in navigation, giving great-circle (shortest) distances between two points on a sphere from their longitudes and latitudes to compute the nearest ESC as shown follow:

$R$ = radius of the earth (average = 6,371 km)

$$\Delta lat = lat_2 - lat_2 \quad (1)$$
$$\Delta long = long_2 - long_2 \quad (2)$$

From equation 1. and 2. obtained the difference between longitude and latitude of the starting point and destination point

$$a = sin^2\left(\frac{\Delta lat}{2}\right) + cos(lat_1).cos(lat_2).sin^2\left(\frac{\Delta long}{2}\right) \quad (3)$$
$$c = 2.atan2(\sqrt{a}.\sqrt{(1-a)}) \quad (4)$$
$$d = R.c \quad (5)$$

d is the distance of two points on the earth's surface

The main steps of the web service are explained in Fig. 8.

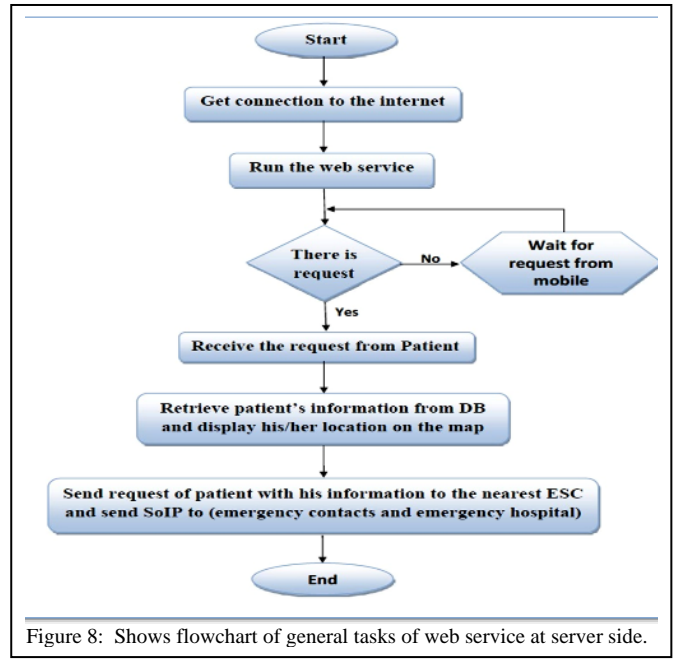

Figure 8: Shows flowchart of general tasks of web service at server side.

### C. The ESC:

Last part is the ESCs, which contains ambulances provided with all required tools for succoring the patients. Fig. 9 shows the main tasks at this part of proposed system.

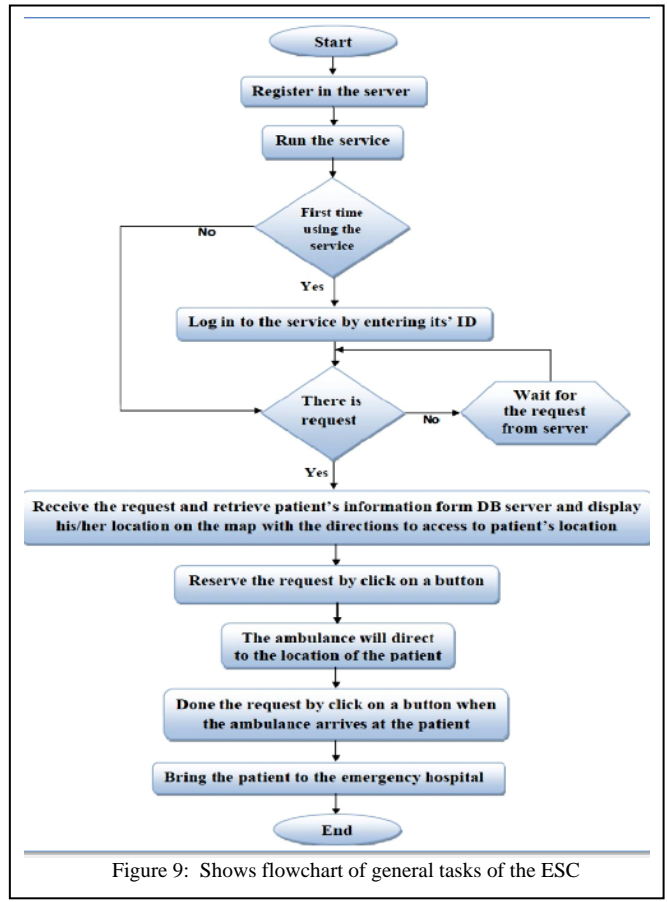

Figure 9: Shows flowchart of general tasks of the ESC

## IV. IMPLEMENT THE PROPOSED SUCCORING SYSTEM

Implement the proposed succoring system for patients via web service on Windows phone (which is friendly with Microsoft packages and cheaper than other devices [1]); this system is called tracking service achieved by using C# language in visual studio 2010 and based on the WCF and WPF technologies as follow:

### A. Technologies Required for Proposed System

There are several important technologies that must be available to design this system:

1) Microsoft visual studio 2010 Service pack 1.
2) Microsoft SQL server 2008 R: The database in this system called (Tracking database) which is a relational database contains 4 tables (a. Registration, b. ESC, c. New_Request and d. Request_Info) as shown in Fig 10.
3) WCF.
4) WPF.
5) Google Maps API V3.1.
6) C# language.
7) Language-Integrated Query (LINQ): the data retrieved from *Tracking* database at web service by using this LINQ.
8) Silver light for windows phone.
9) Windows phone SDK7.1

a) Registration table

b) ESC table

c) New_Request table

d) Request_Info table

Figure 10: Structure of tables at *Tracking* database

### B. From View the Three Parts of Proposed System

1) *Client side:* The important tasks in this side are:
   a) *Registration Steps on the mobile application:* At the first time of using mobile application the patient should be registered in the database which is resided in the server by using the developed application on his/her mobile phone and filling all the required information fields as shown in Fig. 11. When press on the register button, the filled information will be saved in the database of server then the Help page of mobile application will be loaded.

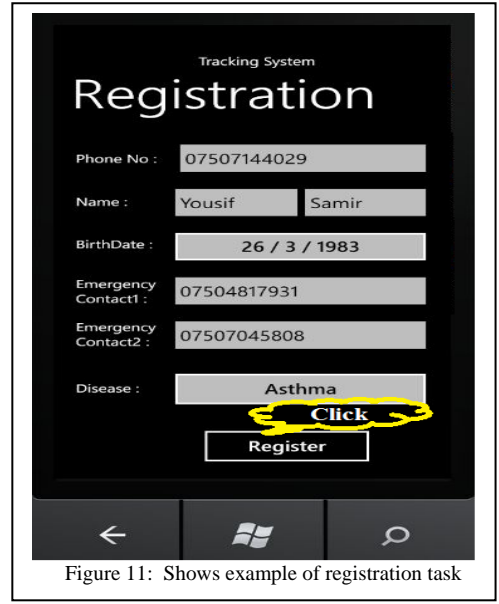

Figure 11: Shows example of registration task

b) *Help at emergency case:* Help facility contains (latitude, longitude and help button). When the patient needed help will click on help button, to send request to the server. The request contain patient's (ID, latitude and longitude, which are obtained their values from GPS satellites automatically because the client used a built-in GPS). Fig. 12 shows GUI of the service at emergency case.

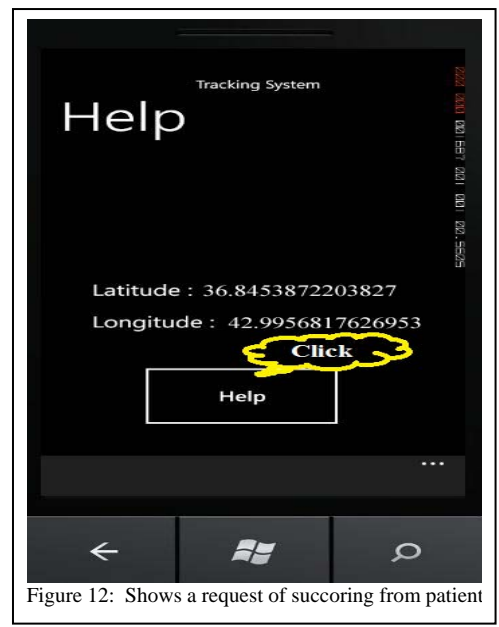

Figure 12: Shows a request of succoring from patient

2) *Server Side:* The server working automatically, when receive the help request from patient's mobile and send it to the nearest ESC, and send SMS to the emergency contacts of the patient when he/she needs help. The server in this system contains two important parts*:*

a) *The web service:* All operations related with this side reveals in the main interface of it, When the client (patient) send the help request from their mobile it will be stored in the database and appeared in the list of request (with red color) which is in the main interface of the web service, the red color of the request means that the request is new ( not handled by the ESCs) and will be converted to black color when it be handled. At the beginning the locations of all requests that are in the list will be displayed on the Google map as multiple markers as shown in Fig. 13.

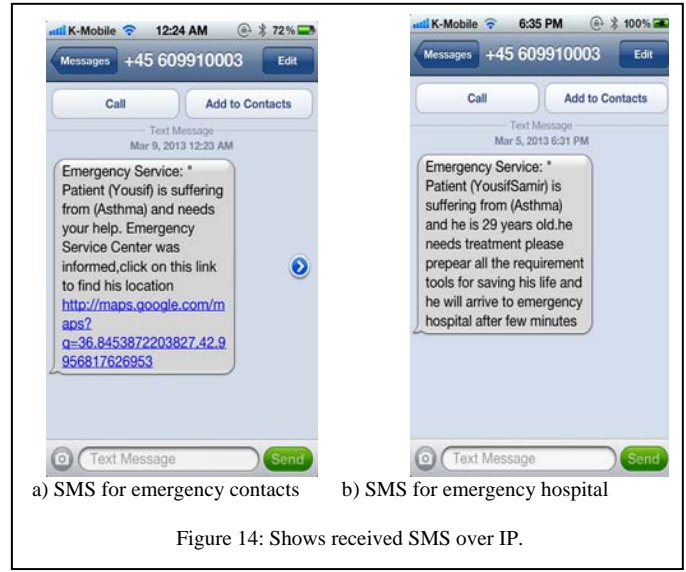

a) SMS for emergency contacts    b) SMS for emergency hospital

Figure 14: Shows received SMS over IP.

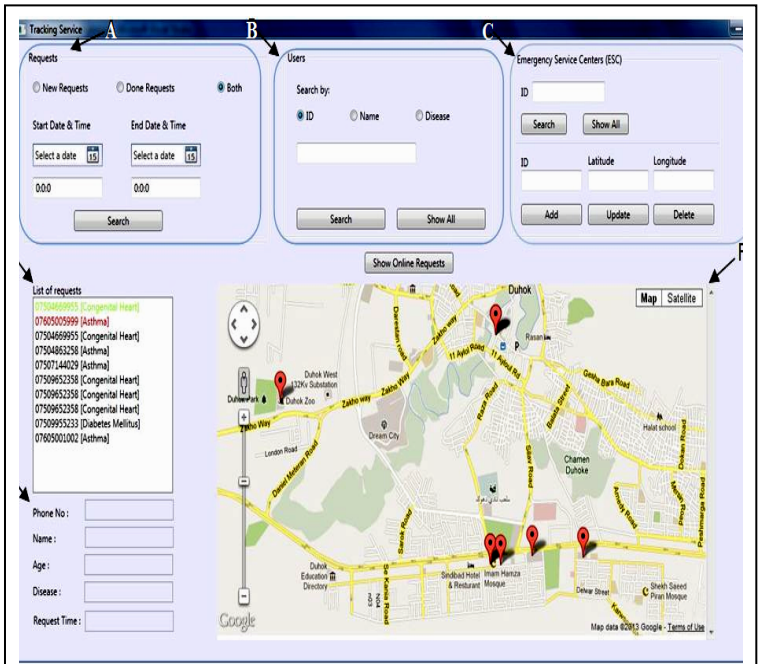

Figure 13: Shows a GUI of web service with multiple markers as follow:

**A.** In this part the server can display the new and handled requests in the list with their locations.
**B.** Display the Information of the patients that are registered in the data base and update it.
**C.** The server can display the information of succoring facilities (e.g. ambulances) that are servicing in the system and edit it.
**D.** This list contains the requests that are arrived to the server form the patients.
**E.** Information of the patients display in this part.
**F.** Display the location of the patients on Google map (map or satellite) view.

When the web service in the server receive the request from the patient's mobile it will send the request to the nearest ESC (using Haversine equation) from the patient and at the same time it will send SMS to the emergency contacts of the patient to be informed that the patient needs help from them, the SMS contain the (name and location) of the patient as shows in Fig 14a, also it send SMS to emergency hospital to prepare all required tools for saving the patient's life as shows in Fig 14b .

When click on one request in the list that will be retrieved their information from the database by using LINQ and display under the list of requests with their location will be plotted on the Google map as single marker as shows in Fig 15. Moreover, the address of the patient that he/she sent the help request from the mobile will appear in the info window.

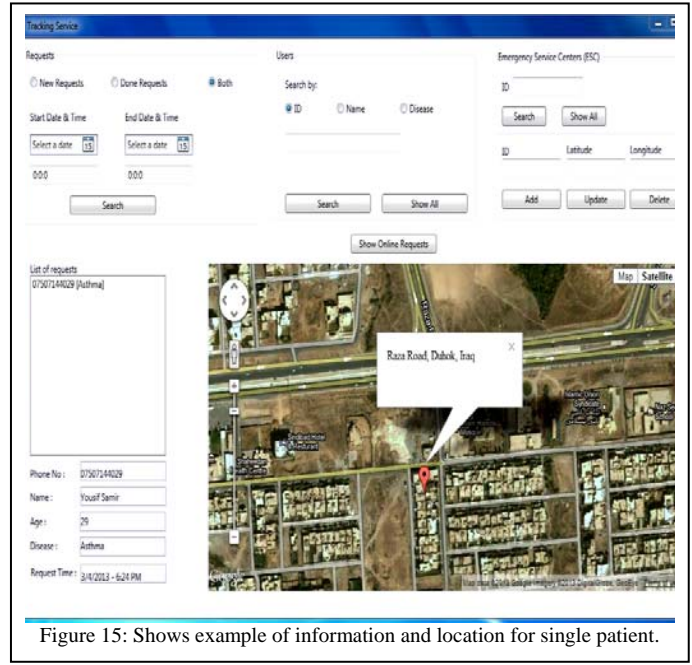

Figure 15: Shows example of information and location for single patient.

b) *The Tracking database*: which contained 4 tables their structures shown in Fig. 10, and the the content of those tables as follow:

  i. *Registration Table:* This table conation information of all registered patients in the database as shown in Fig16.

Figure 16: Shows Registration table in data base.

*ii. ESCs table:* This table contains information about all registered ambulances in the succoring system as shown in Fig 17.

Figure 17: Shows ESCs table in data base.

*iii. New_Requests table:* This table contains information about the new requests (not treated by ESC) that comes from patients as shows in Fig 18.

Figure 18: Shows New_Request table in data base.

*iv. Request_Info:* This table conation information about all requests that been done by ESC as shows in Fig 19.

Figure 19: Shows Request_Info table of data base.

3) *Emergency Service Center:* The service shown at the ESC, , this service should be working 24 hours at day to receive the request from the server, the main interface of ESC shown in Fig. 20.

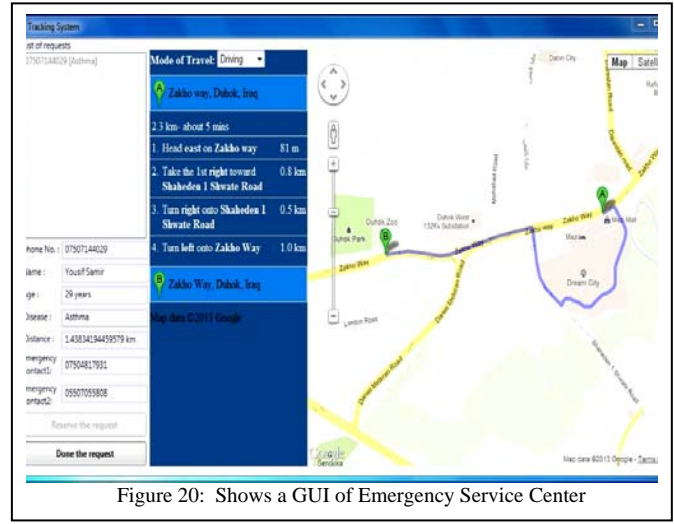

Figure 20: Shows a GUI of Emergency Service Center

From the client (patient's mobile) web service will send it to the nearest ESC from the patient with alarm to inform the workers at ESC there is a new help request. The system contains more than one ESC, which distributed due to the area that used this system, each ESC contains more than one ambulance to serve as facility succoring for the patients when they need help.

*C. Discuss the Results*

Comparing the results of implementing the proposed system with other (related) systems, shown in Table 2.

TABLE II. COMPARING A PROPOSED SYSTEM WITH OTHER SYSTEMS

| Features | Proposed system | Ayad [1] | Ruchika [5] | Khondker[6] |
|---|---|---|---|---|
| Data transferring between client and server | WCF technology (Request & reply) via Internet | SMS via GPRS | GPRS | SMS via GPRS |
| Data transferring continuity | No, just when the client (patient) needs help, but the server works 24 hours a day | Like the proposed system | Yes | Yes |
| The need for dedicated GPS receiver | No , it uses mobile with build in GPS | Like the proposed system | No | Yes |
| The speed of data transferring | WCF provide rate of one request per sec | One request per 10-15 sec | One request in more than 15 sec | One request in more than 15 sec |
| The need to separate the server form database | No, because it used the same server for the web service & database | Like proposed system | Yes | Yes |
| Sending the client request to the nearest emergency service | Yes, display the location of the client (patient) on Google map and send his/her request to nearest ESC to succor patient at shortest time | No, just display the location of client/patient on Google map and send request to ESC without caring to distance | No, just display the present location of the tracked object | No, just display the present location of the object and past history of its movement on the map |
| The supported users | Unlimited, i.e. for all type of patients | Limited, i.e. for Children only | Unlimited | Unlimited |
| Displaying locations of multiple clients (multiple markers) on Google map | Yes | Yes | No | No |

## V. Conclusion and Future Work

The important conclusions obtained from implementing this system are:

1. The using of mobile GIS based on WCF provided a satisfying higher productivity system and this is supported by request and response rate. This response is the highest comparing with other systems.

2. The proposed system achieved higher productivity by selecting the nearest ESC to shorten the time requiring for reaching the patient.

3. The proposed system covers all the aims and objectives, which are required from that system as shown in table 3.

4. Using the WCF technology reduced the expenses comparing to the other systems because of sending the request and receiving the reply done through Internet rather than SMS mode.

The mobile application in the proposed system is developed based on Windows phone, as future work, it can be developed for other cell phones, like iPhone and Blackberry, which used different operation systems, such as iOS and Android.

## References

[1] Ayad Ghany Ismaeel, "An Emergency System for Succoring Children using Mobil GIS", paper appears in (IJACSA) International Journal of Advanced Computer Science and Applications, Vol. 3, No. 9, pages 218-223 2012.

[2] [28][7] Nicki R., Brian R. and Stuart H.,"Davidson's principles and practice of medicine 21st Edition", ISBN-13: 978-0-7020-3085-7, 2010.

[3] Khondker Sh., Mashiur R., Abul L.,M Abdur R., Tanzil R. and M Mahbubur R., "Cost Effective GPS-GPRS Based Object Tracking System",IMECS 2009 Vol I, March 18 - 20, 2009, Hong Kong.

[4] Katina Michael, Andrew McNamee, and MG Michael,"The Emerging Ethics of Humancentric GPS Tracking and Monitoring", Faculty of Informatics, University of Wollongong, Australia [2006], http://ro.uow.edu.au/infopapers/385.

[5] Ruchika Gupta and BVR Reddy, " GPS and GPRS Based Cost Effective Human Tracking System Using Mobile Phones ", VIEWPOINT -June 2011.

[6] Khondker Sh., Mashiur R., Abul L.,M Abdur R., Tanzil R. and M Mahbubur R., "Cost Effective GPS-GPRS Based Object Tracking System",IMECS 2009 Vol I, March 18 - 20, 2009, Hong Kong.

[7] Franklin D. and Jen H.Glendora ,"Medical emergency alert system and method",US 2009/0322513 A1,Dec.31,2009

[8] Braian M.B, Michael C.B, and Nicolas A.N., "panic buttom phone",US006044257A, Mar.28,2000.

[9] Wiley Publishing, Inc. , " Beginning Visual C# 2010" , ISBN: 978-0-470-50226-6, Manufactured in the United States of America 10987654321, 2010.

[10] Matthew MacDonald" Pro WPF: Windows Presentation Foundation in .NET 3.0 " ISBN-13 (pbk): 978-1-59059-782-8, ISBN-10 (pbk): 1-59059-782-6,2007.

[11] Chong-shan Ran, Ning Li ," Design and Implementation of Communication Model Based on Silverlight and WCF in EAM System ", 2010 IEEE

[12] Wei Zhang , Guixue Cheng ,"A Service-Oriented Distributed Framework—WCF", 2009 International Conference on Web Information Systems and Mining, 2009 IEEE.

[13] Markus Stopper and Bernd Gastermann, "Service-oriented Communication Concept based on WCF.NET for Industrial Applications",IMECS 2010,March 17-19,2010, Hong Kong.

[14] Ming-Hsiang Tsou, "Integrated Mobile GIS and Wireless Internet Map Servers for Environmental Monitoring and Management ", *Cartography and Geographic Information Science, Vol. 31, No. 3, 2004, pp. 153-165*

[15] Jiangfeng S., Hongdan J., Weixiong X., Kai X., Jingguang C.," Urban Construction Archive Submission System with WPF Technology", 2009.

[16] Roshan D., Kariyappa B., Santhosh K., Dr. M. Uttara K.," Protocol Implementation for Short Message Service over IP" , ICIIS 2011, Aug. 16-19, 2011, Sri Lanka.

[17] Jane D., Roland B., Elsa J., Elsa J., David F.,"Dynamic and Mobile GIS", ISBN 0‑8493‑9092‑3, G70.212.D96, 2007.

[18] Ayad Ghany Ismaeel, " Effective System for Pregnant Women using Mobile GIS ", paper appears in International Journal of Computer Applications (0975 – 8887) Volume 64– No.11, pages 1-7 February 2013.

AUTHORS PROFILE

**Ayad Ghany Ismaeel** received MSC in computer science from the National Center of Computers NCC- Institute of Postgraduate Studies, Baghdad-Iraq, and Ph.D. Computer science in qualification of computer and IP network from University of Technology, Baghdad- Iraq.

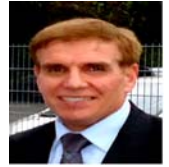

He is professor assistant, at 2003 and currently in department of Information System Engineering in Erbil Technical College- Hawler Polytechnic University (previous FTE Erbil), Iraq. His research interest in mobile, IP networks, Web application, GPS, GIS techniques, distributed systems, bioinformatics and bio-computing. He is senior lecturer in postgraduate of few universities in MSC and Ph.D. courses in computer science and software engineering as well as supervisor of many M.Sc. student additional Higher Diploma in Software Engineering and computer from 2007 till now at Kurdistan-Region, IRAQ.

Ayad Ghany Ismaeel is Editorial Board Member at International Journal of Distributed and Parallel Systems IJDPS http://airccse.org/journal/ijdps/editorial.html, Program Committee Member of conferences related to AIRCC worldwide, and reviewer in IJCNC (which is listed as per the Australian ARC journal ranking http://www.arc.gov.au/era/era_2012/era_journal_list.htm), other journals at AIRCC like IJDPS, IJCSIT etc (http://airccse.org/journal.html), and conferences within AIRCC http://airccse.org/, as well as he adviser and reviewer in multiple national journals. The published papers were in international and national journals about (20) paper.

**Sanaa Enwaya Rizqo** is graduated from computer science of Duhuk University at 2007, and now she is MSc student at research stage in computer science at University of Zakho.

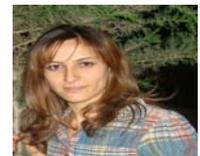

She is working at network department in Computer Institute, Duhok / Iraq. Her researches interest in the fields of Network, IT and Mobile.

Received a certificate in Network Administration from Bit Media e-Learning solutions GmbH & Co KG at the 30th of July 2009.